\documentclass[reprint,amsmath,amssymb,aps,prd,floatfix]{revtex4-2}

\usepackage{graphicx}
\usepackage{dcolumn}
\usepackage{bm}
\usepackage{xcolor}

\begin{document}
\title{Nonequilibrium coherent effects at finite chemical potential}

\author{Amelie Claussen}
 \altaffiliation[amelie.claussen@usm.cl]{}
 \affiliation{
 Universidad Técnica Federico Santa María.\\
 Valparaiso, Chile.
}
\author{Sebastián Mendizabal}
 \altaffiliation[sebastian.mendizabal@usm.cl]{}
\affiliation{
 Universidad Técnica Federico Santa María.\\
 Valparaiso, Chile.
}

\date{\today}

\begin{abstract}

We study a nonequilibrium coherent effect generated by a finite chemical potential in a complex scalar field with a conserved \(U(1)\) charge. The scalar excitation is treated as a probe coupled to an equilibrium thermal reservoir, so the self-energy is an equilibrium kernel and there is no backreaction on the bath. Solving the Schwinger-Keldysh-Kadanoff-Baym equations in the normal phase, when the chemical potential is smaller than the dispersion relation, we keep the particle and antiparticle quasiparticle poles separate. The source-driven inhomogeneous statistical propagator is fixed by the reservoir and relaxes to the usual decoherent equilibrium form. By contrast, the homogeneous solution carries initial-condition memory; finite chemical potential turns this memory into a transient particle-antiparticle interference pattern by splitting the two charge-sector phases. The effect is not a new equilibrium mode, but a phase-sensitive remnant of the initial data that is erased by damping as \(t\to\infty\). We define a normalized interference contrast extracted from the mixed charge-sector terms, illustrate the relaxation using the plasmon damping rate of hot scalar \(\phi^4\) theory, and show that the same normal-phase solution displays the infrared enhancement that precedes Bose-Einstein condensation.

\end{abstract}

\maketitle

\section{\label{sec:level1}Introduction}

The real-time dynamics of quantum fields at finite temperature and nonzero chemical potential is relevant in several areas of high-energy physics. In relativistic heavy-ion collisions, the quark-gluon plasma produced at RHIC and the LHC provides a controlled laboratory for hot QCD matter, while the beam-energy-scan program explores regions of the phase diagram in which the baryon chemical potential is non-negligible~\cite{Busza:2018rrf,Bzdak:2019pkr}. Transport coefficients, equilibration rates, and in-medium spectral functions are intrinsically real-time quantities. At finite density, particle and antiparticle excitations are no longer related by the same effective energy, and this asymmetry can leave transient phase information in nonequilibrium correlation functions before the system has fully relaxed. The charged-scalar model studied below is not intended as a microscopic theory of QCD transport. Its role is narrower and analytically useful: it isolates one generic consequence of a conserved density, namely the real-time splitting between particle and antiparticle propagation. By restricting the analysis to an Abelian $U(1)$ symmetry, we isolate this
coherent dephasing mechanism in a setting where the particle--antiparticle
splitting is analytically transparent. This provides a useful baseline for
more involved finite-density theories with several commuting charge
chemical potentials, such as QCD matter at finite isospin~\cite{Son:2000xc}
or flavor density~\cite{Alford:2007xm}, where additional charge sectors and
mixing effects can enrich the real-time phase structure. The same kinematic issue is present whenever a plasma carries a net
approximately conserved charge, including baryon-rich QCD matter and
cosmological settings in which a baryon asymmetry is represented by an
effective chemical potential or charge imbalance~\cite{Sakharov:1967dj,Cohen:1993nk,Bodeker:2020ghk}.\\

The central question of this paper is how such finite-density phase information appears and disappears. In equilibrium, a chemical potential is usually encoded through shifted frequencies, shifted self-energy arguments, and modified thermal occupation factors~\cite{Loewe:2005df,Loewe:2004zw,Mallik:2009pj,Sasagawa:2011gn,Landsman:1987uw,Chou:1985zz,Weldon:2007zz}. In a two-time nonequilibrium correlator, however, the statistical propagator contains both a source-driven part fixed by the bath and a homogeneous part fixed by initial data \cite{Anisimov:2008dz}. Homogeneous Kadanoff-Baym solutions always carry memory of the initial correlator; the finite-density effect isolated here is more specific. Because the chemical potential \(\mu\) separates the particle and antiparticle phases, that memory contains a transient interference pattern whose oscillation frequency is controlled by the charge-sector energy difference. The chemical potential then acts not only as an equilibrium frequency shift, but also as a dephasing parameter whose effect disappears once damping removes the homogeneous contribution.\\

Finite-density scalar dynamics provides a natural setting for this effect. Relativistic Bose gases and charged scalar sectors exhibit a normal phase bounded by the onset of Bose-Einstein condensation~\cite{Kapusta:1981aa,Haber:1981fg,Haber:1981ts,Bernstein:1990kf,Benson:1991nj,Marko:2014hea}. Closely related scalar dynamics appears in QCD at finite isospin density, where charged pion modes can condense~\cite{Son:2000xc}, and in cosmological settings involving high-occupancy scalar fields. Ultralight or axion-like dark matter is often described by coherent scalar fields with an approximately conserved particle number in the nonrelativistic regime~\cite{Hui:2016ltb}; far-from-equilibrium condensation and relaxation in such systems have been studied as possible mechanisms for forming coherent structures and Bose-star-like objects~\cite{Berges:2012us,Berges:2014xea,Levkov:2018kau,Liebling:2012fv}. These systems are physically different from the probe model analyzed below, but they motivate the same issue: a conserved or effectively conserved density can generate coherent real-time structure before ordinary equilibrium observables are recovered.\\

Conventional imaginary-time methods are powerful for equilibrium thermodynamics, but they do not directly display transient coherence or its decay. Real-time nonequilibrium observables require analytic continuation or a direct closed-time-path formulation~\cite{10.1143/PTP.14.351,lebellac}. The Schwinger-Keldysh-Kadanoff-Baym formalism gives such a direct formulation by evolving spectral and statistical two-point functions on a closed time contour~\cite{Schwinger:1960qe,Keldysh:1964ud,Bakshi:1962dv,Bakshi:1963bn,kadanoff,Danielewicz:1984ww,Chou:1985zz,Landsman:1987uw,Rammer:1986zz,Berges:2004yj}. At finite chemical potential, the shifted time derivative reorganizes the particle and antiparticle sectors. The resulting frequency shifts are familiar in equilibrium, but their imprint on the coherent and decoherent parts of a two-time statistical propagator is the focus of this work.\\

For scalar fields, the same shifted energies also set the boundary of the normal phase. In the notation used below this condition is \(E_{\rm min}(\mu)>0\), where \(E_{\rm min}\) is the smallest shifted quasiparticle energy. When this energy approaches zero, the Bose distribution develops an infrared enhancement and the system approaches Bose-Einstein condensation. Beyond that point the zero mode must be separated from the fluctuating field and treated as a condensate background, as in condensed-phase analyses of relativistic scalar systems~\cite{Nieves:2023yva,Nieves:2024viy,Nieves:2025pdi,Nieves:2024viy}. The coherent effect studied here is therefore a normal-phase phenomenon: it occurs before the condensate is introduced explicitly, while the soft mode is still part of the quasiparticle spectrum.\\

In this work, we study a charged scalar excitation coupled to a thermal reservoir at finite chemical potential. The setup is deliberately simple: the reservoir remains in equilibrium, while the scalar excitation evolves in real time and relaxes toward the bath. In this probe limit there is no backreaction of the scalar excitation on the reservoir, so the self-energy is treated as an equilibrium kernel rather than as a center-time-dependent quantity. Backreaction can modify nonequilibrium spectral functions and requires a separate treatment~\cite{Mendizabal:2016zkk}; here it is left outside the approximation in order to isolate the finite-density coherent transient. This allows us to solve the Kadanoff-Baym equations analytically while keeping the particle and antiparticle poles separate.\\

The main result is the separation of the statistical propagator into two pieces with different coherence properties. The inhomogeneous solution is source-driven by the thermal reservoir and becomes the decoherent fluctuation-dissipation response. The homogeneous solution carries initial-condition memory; at finite \(\mu\), the particle and antiparticle phases split, generating a transient interference pattern whose oscillation frequency is set by the charge-sector energy separation. We quantify this behavior with a normalized contrast extracted from the phase-sensitive part of the homogeneous response. This contrast is a diagnostic of finite-density dephasing, not a replacement for the full statistical propagator. We use the plasmon damping rate of hot scalar \(\phi^4\) theory as a benchmark relaxation scale and show that the same formalism tracks the infrared enhancement associated with the approach to Bose-Einstein condensation.\\

This focus distinguishes the present analysis from earlier nonequilibrium scalar-field studies based on a thermal bath~\cite{Anisimov:2008dz,Mendizabal:2016zkk,Boyanovsky:2004vg,Mendizabal:2025vla}. Those works provide the real-time relaxation framework that we use here, while the present paper keeps the chemical potential as a finite control parameter and isolates the coherent charge-sector transient. Note that, as a difference from \cite{Mendizabal:2025vla}, there is no need for a small chemical potential in order to obtain full results. The shifted Kadanoff-Baym evolution is solved with separate particle and antiparticle poles, the statistical propagator is decomposed into decoherent source-driven and coherent initial-condition pieces, and the homogeneous contribution defines a real-time coherence contrast. The formulation is therefore aimed at finite-density scalar systems in which the magnitude and sign of \(\mu\) are physical control parameters.\\

The paper is organized as follows. In Sec.~II we introduce the Schwinger-Keldysh setup and derive the Kadanoff-Baym equations in the presence of a chemical potential. In Sec.~III we solve the spectral and statistical equations and isolate the finite-density coherent transient. Sec.~IV discusses the plasmon-damping benchmark, the relation to Bose-Einstein condensation, and the numerical illustrations. Sec.~V summarizes the conclusions.


\section{\label{sec:level2}Nonequilibrium dynamics}
We study the system in the Schwinger-Keldysh formalism, where thermal initial conditions and real-time evolution are encoded simultaneously through the density operator 

\begin{equation}
\hat\rho = \frac{\exp\left[-\beta(\hat H-\mu \hat Q)\right]}{Z}, \label{densop}
\end{equation}

Here $\beta=1/T$, $\mu$ is the chemical potential associated with the conserved charge $\hat Q$, and $Z={\rm Tr}\,\exp[-\beta(\hat H-\mu\hat Q)]$ is the partition function.\\

For a complex scalar field with a global \(U(1)\) symmetry, the conserved charge is
\begin{equation}
  Q = \int d^3x\, j^0(x),
\end{equation}
with current
\begin{equation}
  j^\mu(x)
  = i\left[
    \phi^\dagger(x)\partial^\mu\phi(x)
    - \partial^\mu\phi^\dagger(x)\phi(x)
  \right].
\end{equation}

We use this formalism for a probe scalar sector coupled to a reservoir whose state remains thermal and homogeneous. Thus the self-energy is taken as an equilibrium kernel, with no independent dependence on the center time, while the scalar two-point functions may still evolve out of equilibrium because their statistical component carries initial-condition data. This is an equilibrium-reservoir approximation rather than a self-consistent kinetic calculation: it does not include backreaction of the probe on the bath, nor the reorganization of the spectrum that occurs after condensation. Its purpose is to isolate how a conserved-charge chemical potential modifies the real-time relaxation of normal-phase scalar correlations.\\

Let us see how the chemical potential structurally modifies the field dynamics. Following \cite{Weldon:2007zz, Loewe:2005df}, the grand-canonical density operator in Eq. \eqref{densop} dictates that the chemical potential $\mu$ couples to the conserved $U(1)$ charge as a constant background temporal gauge field $A_0=\mu$. So, the Heisenberg time evolution of the charged field operators acquires an intrinsic phase shift. At the level of the action and the resulting equations of motion, this coupling is captured y promoting the ordinary time derivative to a covariant derivative, $\partial_t \to \partial_t - i\mu q$, where $q=1$ for the particle excitation of the fundamental scalar field. Thus, for a complex scalar field in the presence of this chemical potential, the contour Green's function satisfies the shifted Schwinger-Dyson equation
 \begin{align}\label{schwingerdysoneq}
 &((\partial/\partial t_1-i\mu)^2-\nabla^2+m^2)G_C(x_1,x_2)&\nonumber\\&+i\int_Cd^4x'\Pi_C(x_1,x')G_C(x',x_2)=-i\delta_C(x_1-x_2),
 \end{align}
where $\Pi_{\cal C}$ denotes the contour self-energy, $\delta_{\cal C}$ is the contour delta function, and the closed-time contour ${\cal C}$ runs from $-\infty$ to $+\infty$ and back. The chemical potential enters through the shifted time derivative, reflecting the coupling to the conserved $U(1)$ charge.\\

It is convenient to decompose contour correlators into their greater and lesser components, 

\begin{eqnarray}
O^>(x_1,x_2)\equiv \langle \Phi(x_1)\Phi^\dagger(x_2)\rangle,
\\
O^<(x_1,x_2)\equiv \langle \Phi^\dagger(x_2)\Phi(x_1)\rangle,
\end{eqnarray}

and then define the symmetric and antisymmetric combinations
 \begin{eqnarray}\label{plusminusdefinitiosa}
     \mathcal O^+(x_1,x_2)\equiv & \frac{1}{2}\big( \mathcal O^>(x_1,x_2)+\mathcal O^<(x_1,x_2)\big),\\\label{plusminusdefinitiosb}
     \mathcal O^-(x_1,x_2)\equiv &  i\big( \mathcal O^>(x_1,x_2)-\mathcal O^<(x_1,x_2)\big).
 \end{eqnarray}
In what follows, these definitions will be applied both to the propagators $G^\pm$ and to the self-energies $\Pi^\pm$. The antisymmetric correlator $G^-$ is the spectral function, while the symmetric correlator $G^+$ is the statistical propagator.\\

If the thermal reservoir remains in equilibrium, the self-energies are invariant under time translations, $\Pi(x_1,x_2)=\Pi(x_1-x_2),$ and the corresponding equilibrium spectral function depends only on the time difference, $G^-(x_1,x_2)=G^-(x_1-x_2)$ \cite{Anisimov:2008dz}.  Assuming spatial homogeneity and isotropy, and working in momentum space, the Kadanoff-Baym equations reduce to
\begin{align}\label{firstKB}
   ((\partial/\partial t_1-i\mu)^2&+\omega_{\bf k}^2)G^-_{\mu,\bf k}(t_1-t_2)\nonumber\\=-&\int_{t_2}^{t_1}dt'\Pi^-_{\bf k}(t_1-t')G_{\mu,\bf k}^-(t'-t_2),
   \end{align}
for the spectral function, and
   \begin{align}\label{secondKB}
   ((\partial/\partial t_1-i\mu)^2&+\omega_{\bf k}^2)G^+_{\mu,\bf k}(t_1,t_2)\nonumber\\=&- \int_{t_i}^{t_1}dt'\Pi^-_{\bf k}(t_1-t')G^+_{\mu,\bf k}(t',t_2)\nonumber \\
   &+\int_{t_i}^{t_2}dt'\Pi^+_{\bf k}(t_1-t')G^-_{\mu,\bf k}(t'-t_2).
   \end{align}
for the statistical propagator. Here $\omega_{\bf k}^2={\bf k}^2+m^2$ and $t_i$ denotes the initial time.\\

The nonequilibrium character of the system is reflected in the two-time dependence of the statistical propagator. It is therefore useful to introduce the relative and center times,
\[
  y \equiv t_1-t_2,\qquad
  t \equiv \frac{t_1+t_2}{2}.
\]
The variable $y$ describes the separation between the two field insertions, while the center time $t$ tracks the macroscopic evolution of the system. The approach to equilibrium corresponds to the late-time limit $t\to\infty$, in which time-translation invariance is gradually recovered and the fluctuation-dissipation relation assumes its equilibrium form. In the presence of a chemical potential, however, one must treat this limit with care, since the usual frequency shifts can modify the standard form of the Kubo-Martin-Schwinger relation~\cite{Kubo:1957mj,Martin:1959jp}.\\

Equations~\eqref{firstKB} and \eqref{secondKB} provide the starting point for our analysis. In the next section, we solve these equations for a scalar excitation coupled to an equilibrium thermal bath and show that the chemical potential induces qualitatively distinct effects in the spectral and statistical sectors. In particular, we will see that the real-time solutions retain the separate particle and antiparticle contributions that are hidden inside the usual equilibrium frequency-space description.


\section{\label{sec:level3}Scalar correlators below the condensation threshold}

The solution is organized in the quasiparticle regime in which the retarded self-energy produces well separated poles with widths small compared with the corresponding real parts. The chemical potential is restricted only by the normal-phase condition derived below; once a shifted pole energy reaches zero, the zero mode must be treated separately and the present expansion ceases to be a complete description of the system.\\

The spectral function can be obtained directly from Eq.~\eqref{firstKB}. Since it depends only on the relative time $y=t_1-t_2$, the equation can be simplified by changing variables and performing a Laplace transform with respect to $y$. The resulting convolution with the self-energy becomes algebraic, and the inverse transform is written as a Bromwich integral,
\begin{align}
G_{\bf k,\mu}^-(y)
&=
\int_{\mathcal{C_B}}\frac{ds}{2\pi i}
e^{sy}\frac{N_\mu(s)}{D_\mu(s)},
\label{deltaminusbromwich}\\
N_\mu(s)
&=
\dot{G}_{\mu,\bf k}^-(0)
+sG_{\mu,\bf k}^-(0)
-2i\mu G_{\mu,\bf k}^-(0),
\nonumber\\
D_\mu(s)
&=
s^2-2i\mu s+(\omega_{\bf k}^2-\mu^2)
+\tilde{\Pi}_{\bf k}(s).
\nonumber
\end{align}

Here ${\cal C}_B$ denotes the Bromwich contour and $\widetilde\Pi_k(s)$ is the Laplace transform of the self-energy kernel.\\

Note that under $y\to -y$, one must simultaneously account for the change $\mu\to -\mu$. The spectral function satisfies the usual equal-time conditions 
\begin{eqnarray}
G^-_{\bf k,\mu}(0)=0,
\\
\dot{G}_{\bf k,\mu}^-(0)=\partial_y G^-_{\bf k,\mu}(y)\big|_{y=0}=1.
\end{eqnarray}

Using these conditions, one may write the spectral function in terms of its energy representation as

\begin{equation}\label{gimutiempoint}
    G_{\bf k,\mu}^-(y)=i\int\frac{d\omega}{2\pi}e^{-i\omega y}e^{i\mu y} \rho_{\mu,\bf k}(\omega),
\end{equation}
where the spectral density is
\begin{align}
    \rho_{\bf k,\mu}(\omega)
    &=
    \frac{2\,\text{Im}\Pi_{\bf k}^R(\omega+\mu)}
    {{\cal D}_{\bf k,\mu}(\omega)},
    \label{dsdsds}\\
    {\cal D}_{\bf k,\mu}(\omega)
    &=
    [\omega^2-\omega_{\bf k}^2
    -\text{Re}\Pi_{\bf k}^R(\omega+\mu)]^2
    \nonumber\\
    &\quad
    +[\text{Im}\Pi_{\bf k}^R(\omega+\mu)]^2 .
    \nonumber
\end{align}

This expression shows that the spectral function retains the familiar Breit-Wigner structure, with the chemical potential entering through the shifted argument $\omega+\mu$ of the retarded self-energy.\\

To make the pole structure explicit, we now assume an isolated narrow quasiparticle peak in each charge sector. The self-energy entering this pole approximation is the equilibrium reservoir self-energy specified above; the pole parameters are therefore not obtained from a center-time-dependent, backreacting self-consistency problem. Within this approximation, the finite chemical potential is kept in the shifted pole energies throughout the normal phase. We introduce the effective quasiparticle frequencies and widths,

\begin{align}
\Omega_{\bf k}^{\pm}&=\sqrt{\omega_{\bf k}^2+\text{Re}\Pi_{\bf k}^R(\omega_{\bf k}\pm\mu)},\label{omegadef}\\
\Gamma_{\bf k}^{\pm}&=\frac{\text{Im}\Pi_{\bf k}^R(\Omega_{\bf k}^{\pm}\pm\mu)}{\Omega_{\bf k}^{\pm}}\operatorname{sgn}(\Omega_{\bf k}^{\pm}).\label{gammadef}
\end{align}
After the quasiparticle poles have been specified, the shifted energies that enter the thermal weights are
\begin{equation}
  E_{\pm}({\bf k},\mu)=\Omega_{\bf k}^{\pm}\pm\mu .
  \label{shiftedenergies}
\end{equation}
The normal-phase analysis requires
\begin{equation}
  E_{\pm}({\bf k},\mu)>0
  \label{normalphasecondition}
\end{equation}
for all momenta in both charge sectors. Throughout this paper, the \(-\) branch denotes the particle sector, with \(E_-({\bf k},\mu)=\Omega_{\bf k}^{-}-\mu\), while the \(+\) branch denotes the antiparticle sector, with \(E_+({\bf k},\mu)=\Omega_{\bf k}^{+}+\mu\). Thus a positive chemical potential lowers the particle branch, while a negative chemical potential lowers the antiparticle branch. The normal-phase solution is valid only while both shifted pole energies remain positive; the condensation threshold is reached when the minimum shifted energy vanishes.\\

Evaluating the frequency integral by residues then gives
\begin{align}
\label{spectralfunctminusplus}
     G^-_{\bf k,\mu}(y)=\frac{e^{i\mu y}}{i(\Omega_{\bf k}^++\Omega_{\bf k}^-)}\bigg[e^{i\Omega_{\bf k}^+y}e^{-\frac{\Gamma_{\bf k}^+|y|}{2}}
     -e^{-i\Omega_{\bf k}^-y}e^{-\frac{\Gamma_{\bf k}^-|y|}{2}}\bigg].
 \end{align}

Equation~\eqref{spectralfunctminusplus} makes clear that the real-time spectral function contains two oscillatory contributions associated with the two charge sectors. Within the normal phase, the two sectors are characterized by independent frequencies and widths. In the limit $\mu\rightarrow0$ one regains the normal oscillatory one mode spectral function.
\begin{equation}
   G^-_{\bf k,\mu=0}(y)=\frac{1}{\Omega_{\bf k}} \sin(\Omega_{\bf k}y)e^{-\Gamma_{\bf k}y/2}.
\end{equation}

We now turn to the statistical propagator. Following the standard treatment of the Kadanoff-Baym equations, the solution of Eq.~\eqref{secondKB} can be decomposed into an inhomogeneous part and a homogeneous part, 

\begin{equation}
G^+_{\bf k,\mu}(t_1,t_2)
=
G^I_{\bf k,\mu}(t_1,t_2)+G^H_{\bf k,\mu}(t_1,t_2).
\end{equation}

The inhomogeneous contribution is fully determined by the spectral function and can be written as

\begin{align}\label{gmemnonequichem}
    & G_{\bf k,\mu}^{I}(t_1,t_2)=\int\frac{d\omega}{2\pi}\bigg(\int_{t_i}^{t_1}dt'G_{\bf k,\mu}^-(t_1-t')e^{i\omega(t_1-t')}\bigg)\nonumber\\ 
     &\times\Pi_{\bf k}^+(\omega)\bigg(\int_{t_i}^{t_2}dt''G_{\bf k,\mu}^-(t''-t_2)e^{-i\omega(t_2-t'')}\bigg)e^{-i\omega(t_1-t_2)},
 \end{align}

For simplicity, we choose the initial time to be $t_i=0$. In this expression, the chemical potential enters through the spectral functions inside the time integrals. Since each spectral function contains contributions from both charge sectors, the inhomogeneous solution keeps the full finite-density charge-sector structure.\\

Inserting (\ref{spectralfunctminusplus}) into (\ref{gmemnonequichem}) and shifting the integration variable according to $\omega \to \omega-\mu$, one obtains

\begin{align}
     G_{\bf k,\mu}^{I}(t_1,t_2)
     &=
     \int\frac{d\omega}{2\pi}
     \frac{\Pi_{\bf k}^+(\omega-\mu)e^{-i\omega(t_1-t_2)}
     e^{i\mu(t_1-t_2)}}{[i(\Omega_{\bf k}^++\Omega_{\bf k}^-)]^2}
     \nonumber\\
     &\times
     \left[{\cal A}_+(t_1,\omega)-{\cal A}_-(t_1,\omega)\right]
     \nonumber\\
     &\times
     \left[{\cal B}_+(t_2,\omega)-{\cal B}_-(t_2,\omega)\right],
     \label{fullgplusbeforeomega}
\end{align}
where
\begin{align}
{\cal A}_+(t,\omega)
&=
\frac{e^{i(\omega+\Omega_{\bf k}^++i\Gamma_{\bf k}^+/2)t}-1}
{i(\omega+\Omega_{\bf k}^++i\Gamma_{\bf k}^+/2)},\nonumber\\
{\cal A}_-(t,\omega)
&=
\frac{e^{i(\omega-\Omega_{\bf k}^-+i\Gamma_{\bf k}^-/2)t}-1}
{i(\omega-\Omega_{\bf k}^-+i\Gamma_{\bf k}^-/2)},\nonumber\\
{\cal B}_+(t,\omega)
&=
\frac{e^{i(-\omega-\Omega_{\bf k}^++i\Gamma_{\bf k}^+/2)t}-1}
{i(-\omega-\Omega_{\bf k}^++i\Gamma_{\bf k}^+/2)},\nonumber\\
{\cal B}_-(t,\omega)
&=
\frac{e^{i(-\omega+\Omega_{\bf k}^-+i\Gamma_{\bf k}^-/2)t}-1}
{i(-\omega+\Omega_{\bf k}^-+i\Gamma_{\bf k}^-/2)}.
\end{align}

Assuming the reservoir remains in thermal equilibrium, the symmetric self-energy may be related to the retarded one through the Kubo-Martin-Schwinger relation~\cite{Kubo:1957mj,Martin:1959jp}, 

\begin{equation}
\Pi_{\bf k}^+(\omega)
=
\coth\!\left(\frac{\beta\omega}{2}\right)
{\rm Im}\,\Pi_{\bf k}^R(\omega).
\end{equation}
For a charged excitation, the KMS factor is evaluated on the energy conjugate to the grand-canonical evolution generated by \(H-\mu Q\)~\cite{Weldon:2007zz}. Equivalently, a mode with charge \(q\) and quasiparticle energy \(\Omega_{\bf k}\) carries the Bose weight \(n_B(\Omega_{\bf k}-q\mu)\). With the conventions of Eq.~\eqref{shiftedenergies}, the particle pole has \(q=+1\) and therefore samples \(E_-({\bf k},\mu)=\Omega_{\bf k}^--\mu\), while the antiparticle pole has \(q=-1\) and samples \(E_+({\bf k},\mu)=\Omega_{\bf k}^++\mu\). Thus the residues of the two pole contributions carry
\begin{equation}
  C_\beta^\pm({\bf k},\mu)
  =
  1+2n_B(E_\pm({\bf k},\mu))
  =
  \coth\!\left[\frac{\beta E_\pm({\bf k},\mu)}{2}\right].
  \label{shiftedkmsweights}
\end{equation}
This is the sense in which the chemical potential must be retained in the shifted KMS argument: the weights are assigned to \(E_\pm\), not to an unshifted quasiparticle frequency alone.\\

Keeping only the leading contributions in the narrow-width expansion, namely terms of order $1/(\Omega_{\bf k}^{\pm})^2$, one obtains the inhomogeneous solution in terms of center and relative times $t=(t_1+t_2)/2$ and $y=t_1-t_2$ respectively,

\begin{align}\label{resultplus}
    G_{\bf k,\mu}^I(t,y)=\frac{e^{i\mu y}}{(\Omega_{\bf k}^++\Omega_{\bf k}^-)^2}\nonumber\\
    \times\Bigg[\Omega_{\bf k}^+C_{\beta}^+F^+(y,t)e^{i\Omega_{\bf k}^+y}
    +&\Omega_{\bf k}^-C_{\beta}^-F^-(y,t)e^{-i\Omega_{\bf k}^-y}\Bigg],
\end{align}
with 
\begin{equation}
F^{\pm}(y,t)=e^{-\Gamma_{\bf k}^{\pm}t}-e^{-\Gamma_{\bf k}^{\pm}|y|/2}
\end{equation}
and \(C_{\beta}^{\pm}({\bf k},\mu)\) given by Eq.~\eqref{shiftedkmsweights}.\\

Equation~\eqref{resultplus} shows that the inhomogeneous part of the statistical propagator is fixed by the reservoir. It keeps the shifted particle and antiparticle poles, and the chemical potential modifies the oscillation frequencies and thermal weights. At late center time \(t\to\infty\), the terms proportional to \(e^{-\Gamma_{\bf k}^\pm t}\) vanish and \(G_{\bf k,\mu}^I\) becomes the stationary, source-driven contribution required by the fluctuation-dissipation relation. In this sense the inhomogeneous solution is decoherent: it contains the bath-imposed charge-sector spectrum but no independent memory of the initial particle-antiparticle phase. Thus the chemical potential is not an overall phase multiplying the zero-density propagator. It changes the locations and widths of the particle and antiparticle poles separately, so the real-time solution can display different oscillation frequencies and decay rates in the two charge sectors.\\

The homogeneous contribution contains the information associated with the initial conditions. It can be written as
\begin{align}
    G_{\bf k,\mu}^{H}(t_1,t_2)=A_{\mu}(t_2)\dot{G}_{\bf k,\mu}^-(t_1)+B_{\mu}(t_2)G_{\bf k,\mu}^-(t_1).
\end{align}
where the functions $A_\mu$ and $B_\mu$ are fixed by the initial data. Using the symmetry relation $G^-_{\bf k,-\mu}(-y) = - G^-_{\bf k,\mu}(y)$, one finds

\begin{align}
    G_{\bf k,\mu}^{H}(t_1,t_2)&=G_{in}^+\dot{G}_{\bf k,\mu}^-(t_1)\dot{G}_{\bf k,-\mu}^-(t_2)\nonumber\\
    &+\ddot{G}_{in}^+G_{\bf k,\mu}^-(t_1)G_{\bf k,-\mu}^-(t_2)\nonumber\\
    &+\dot{G}_{in}^+\dot{G}_{\bf k,\mu}^-(t_1)G_{\bf k,-\mu}^-(t_2)\nonumber\\
    &+\dot{G}_{in}^+G_{\bf k,\mu}^-(t_1)\dot{G}_{\bf k,-\mu}^-(t_2),
\end{align}
 Here $G_{in}^+$, $\dot{G}_{in}^+$ and $\ddot{G}_{in}^+$ are the initial conditions of the statistical propagator. \\

The homogeneous contribution is the part that carries the coherent nonequilibrium effect. It transports the initial data through the finite-density quasiparticle modes. At nonzero chemical potential, the particle and antiparticle sectors no longer evolve with the same effective phase: the spectral function separates into damped modes with frequencies and widths \(\Omega_{\bf k}^-\), \(\Gamma_{\bf k}^-\) for particles and \(\Omega_{\bf k}^+\), \(\Gamma_{\bf k}^+\) for antiparticles. Products such as \(G^-_{\bf k,\mu}(t_1)G^-_{\bf k,-\mu}(t_2)\) and their time derivatives generate mixed charge-sector terms whose phases are determined by combinations of the two quasiparticle frequencies. The magnitude of this coherent signal is set by \(G_{in}^+\), \(\dot G_{in}^+\), and \(\ddot G_{in}^+\): larger initial correlations produce a larger transient, while the exponential factors in the spectral functions suppress it at late time.\\

To make this statement operational, we can focus on the mixed equal-time terms in the homogeneous response, namely the terms with relative particle-antiparticle phases. In the narrow-width limit these phases are controlled by
\begin{equation}
  \Delta E_{\bf k}(\mu)
  =
  E_+({\bf k},\mu)-E_-({\bf k},\mu),
  \qquad
  \bar\Gamma_{\bf k}
  =
  \frac{\Gamma_{\bf k}^{+}+\Gamma_{\bf k}^{-}}{2}.
\end{equation}
A normalized interference contrast can then be read from this phase-sensitive part of the homogeneous propagator,
\begin{equation}
  {\cal C}_{\bf k}(t;\mu)
  =
  e^{-\bar\Gamma_{\bf k}t}
  \cos\!\left[\Delta E_{\bf k}(\mu)t\right],
  \label{coherencecontrast}
\end{equation}
up to an overall normalization fixed by the initial correlator. Equation~\eqref{coherencecontrast} is not the full homogeneous propagator; it is a compact diagnostic obtained by projecting onto the mixed particle-antiparticle phases. This distinction is important because ordinary homogeneous Kadanoff-Baym evolution already contains initial-condition memory. The finite-density content of the present result is the \(\mu\)-dependent phase splitting in that memory, encoded by \(\Delta E_{\bf k}(\mu)\). At \(\mu=0\), the two shifted charge-sector energies coincide in a charge-symmetric medium and the contrast decays only through damping. Residual wiggles that may remain in a numerical equal-time signal at \(\mu=0\) are not the finite-density coherent effect; in the narrow-width expansion they are small contributions of order \(\Gamma_{\bf k}/(\Omega_{\bf k}^{\pm})^2\) from the damped quasiparticle kernels. For \(\mu\neq0\), the finite separation \(\Delta E_{\bf k}\) produces genuine particle-antiparticle dephasing. Larger \(|\mu|\) gives faster oscillations because the two phases separate more rapidly, but it also leads to faster loss of the coherent signal whenever the shifted damping rates increase. Thus the chemical potential enhances the visibility of the transient oscillatory structure while also accelerating its disappearance. In the limit \(t\to\infty\), all homogeneous terms die away and only the decoherent inhomogeneous, equilibrium-reservoir contribution remains.


\section{\label{sec:discussion}Discussion}

\subsection{\label{sec:damping}Damping scale and numerical illustration}

For the explicit damping scale used in the numerical discussion, we take as reference the plasmon damping rate computed by Aarts and Smit in hot classical scalar \(\phi^4\) theory~\cite{Aarts:1996qi}. In their analysis of the real-time two-point function at finite temperature, the leading high-temperature damping rate is obtained from the large-time behavior of the correlation function. We do not assume that this rate is a complete self-energy for every microscopic realization of the reservoir. Rather, it supplies a transparent benchmark dissipative scale that can be assigned to each shifted charge sector,
\begin{equation}
  \Gamma_{\rm AS}^{\pm}
  = \frac{\lambda^2 T^2}{1536\pi E_\pm}.
  \label{eq:aswidth}
\end{equation}
In the limit \(\mu\to0\), one has $\Omega_{\bf k}^{+}=\Omega_{\bf k}^{-}$ and the original Aarts-Smit expression is recovered. Here this width sets a benchmark relaxation scale for the quasiparticle propagators while the system remains in the normal phase. Thus the damping rate fixes the dissipative time scale of the relaxation, while the finite-density dependence follows from the separate particle and antiparticle poles. A different microscopic reservoir would change the numerical widths, but not the pole separation, the thermal enhancement or the dephasing structure in Eq.~\eqref{coherencecontrast}.\\

For the scalar $\phi^4$ theory, we use the charge-symmetric quasiparticle benchmark \(\Omega_{\bf k}^{+}=\Omega_{\bf k}^{-}\); for the plotted \(k=0\) mode this common frequency is denoted by \(\Omega_0\). The coherent finite-density transient is especially visible in the homogeneous solution because this piece transports the initial data without being fixed by the reservoir source. Figure~\ref{fig0} shows an equal-time homogeneous response \(G^H_{\bf k,\mu}(t)\) for three chemical potentials. The plotted quantities are dimensionless: the panels use \(\mu/\Omega_0=0.9,0.6,0.2\), while the curves use \(G_{in}^+\Omega_0=1,5,10,20\), so the initial condition is measured in units of \(1/\Omega_0\). Increasing the initial correlation amplifies the homogeneous transient, while changing \(\mu\) modifies the oscillation pattern and damping through the separated particle and antiparticle poles.\\

\begin{center}
    \centering
    \includegraphics[width=90mm]{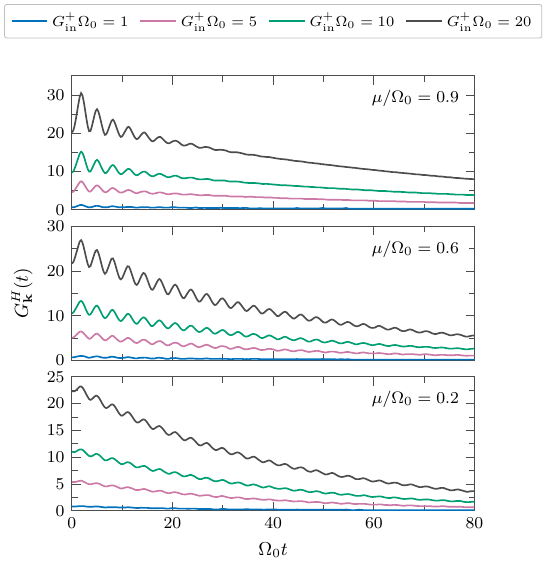} 
    \refstepcounter{figure}\label{fig0}
    \smallskip
    \parbox{0.95\columnwidth}{\small\textbf{FIG.~\thefigure.} Homogeneous response \(G^H_{\bf k}(t)\) with $y=0$, $\lambda=0.0005$ and $T=7\Omega_0$. From top to bottom, \(\mu/\Omega_0=0.9,0.6,0.2\). In each panel, the curves use \(G_{in}^+\Omega_0=1,5,10,20\), corresponding to initial conditions in units of \(1/\Omega_0\).}
\end{center}

A complementary slice keeps the initial condition fixed and varies only the chemical potential. Figure~\ref{fig:severalmu} shows the homogeneous response for fixed initial condition and different chemical potential, \(G_{in}^+\Omega_0=1\), corresponding to \(G_{in}^+=1/\Omega_0\), and for several dimensionless chemical potentials \(\mu/\Omega_0\). The larger chemical potentials produce a more visible oscillatory transient, while the damping envelope still drives the homogeneous contribution toward zero at late times.\\

\begin{center}
    \centering
    \includegraphics[width=\columnwidth]{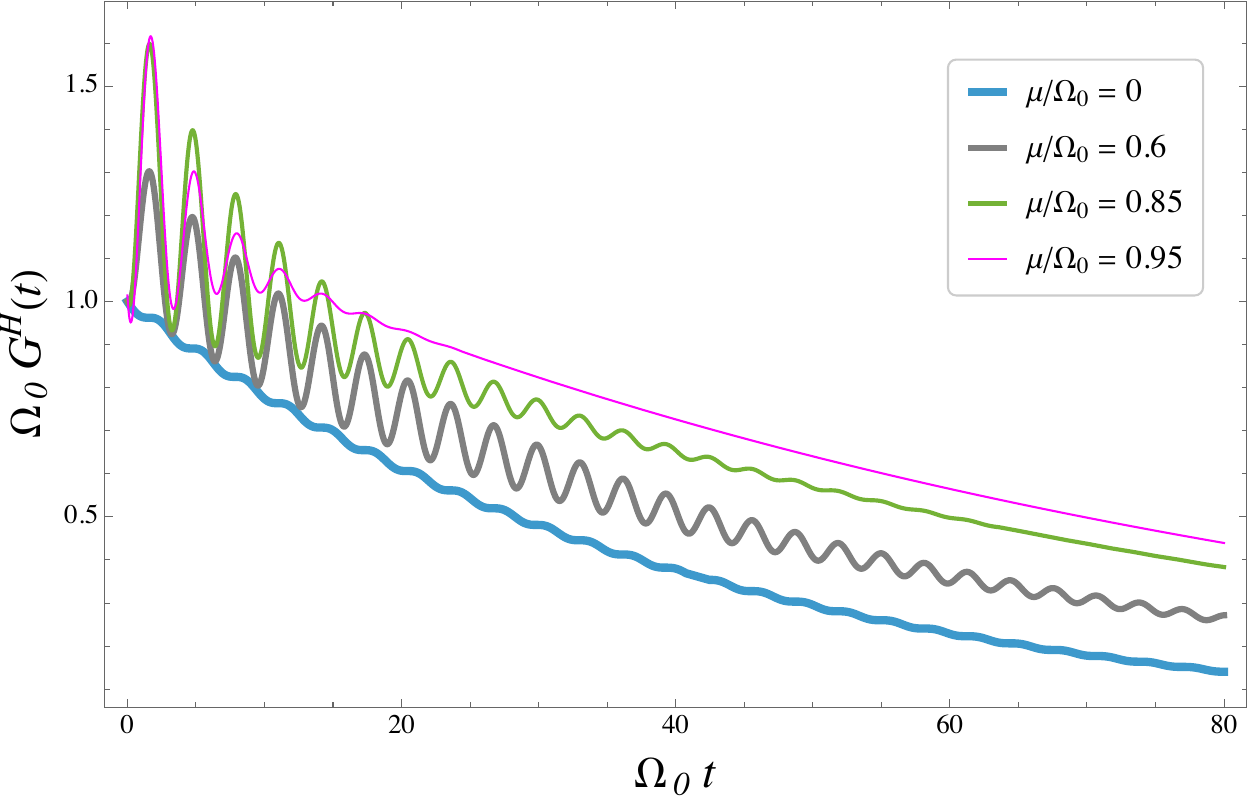}
    \refstepcounter{figure}\label{fig:severalmu}
    \smallskip
    \parbox{0.95\columnwidth}{\small\textbf{FIG.~\thefigure.} Homogeneous response \(G^H_{\bf k}(t)\) for fixed \(G_{in}^+\Omega_0=1\), i.e. \(G_{in}^+=1/\Omega_0\), and several dimensionless chemical potentials \(\mu/\Omega_0=0,0.6,0.85,0.95\) at $y=0$, $\lambda=0.0005$ and $T=7\Omega_0$.}
\end{center}

\begin{center}
    \includegraphics[width=\columnwidth]{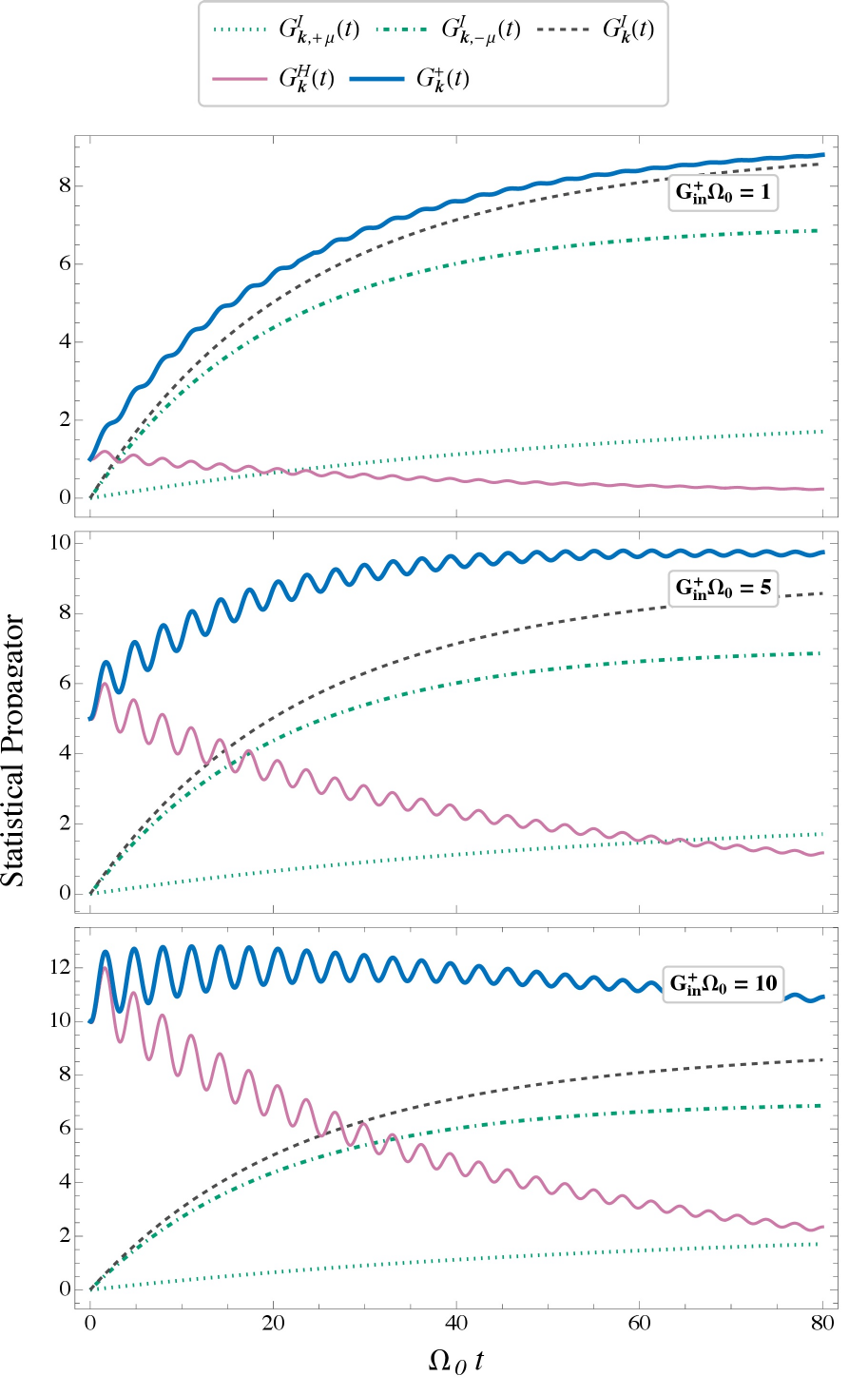}
    \refstepcounter{figure}\label{fig1}
    \smallskip
    \parbox{0.95\columnwidth}{\small\textbf{FIG.~\thefigure.} Finite-density statistical response at \(\mu/\Omega_0=0.5\) and $y=0$, $\lambda=0.0005$ and $T=7\Omega_0$. From top to bottom, the panels use \(G_{in}^+\Omega_0=1,5,10\). The curves show the total response, the inhomogeneous contribution \(G^I_{\bf k}(t)\), its shifted charge-sector pieces \(G^I_{\bf k,+\mu}(t)\) and \(G^I_{\bf k,-\mu}(t)\), and the homogeneous remnant \(G^H_{\bf k}(t)\).}
\end{center}

The full statistical propagator is obtained by adding the source-driven inhomogeneous contribution to the homogeneous piece. The inhomogeneous contribution inherits the two-sector structure of the spectral function, but it is fixed by the reservoir and becomes the decoherent equilibrium response. The homogeneous contribution keeps the initial-condition-dependent coherent part and is gradually suppressed by damping as the statistical propagator relaxes. Figure~\ref{fig1} shows this decomposition at fixed chemical potential \(\mu/\Omega_0=0.5\). The three panels correspond to \(G_{in}^+\Omega_0=1,5,10\), so they illustrate how increasing the initial condition enhances the homogeneous remnant before it decays into the reservoir-dominated response.\\

\subsection{\label{sec:bec}Approach to Bose-Einstein condensation}

Before discussing the thermodynamics of the normal phase, it is useful to
emphasize the dual role played by the chemical potential. In the preceding
analysis, \(\mu\) acted primarily as a dynamical control parameter: increasing
\(|\mu|\) enlarges the particle--antiparticle phase separation and therefore
raises the frequency of the transient coherent oscillations. The same
parameter, however, also controls the thermodynamic stability of the scalar
plasma. The shifted quasiparticle energies move in opposite directions; for
positive \(\mu\), the particle branch \(E_-\) is driven downward, whereas for
negative \(\mu\), the antiparticle branch \(E_+\) is driven downward. Thus,
the chemical potential simultaneously enhances real-time dephasing and pushes
one charge sector toward the infrared. As a result, the dynamical enhancement
of the nonequilibrium coherent transient is bounded by the normal-phase
condition that precedes Bose--Einstein condensation.\\

\begin{center}
    \centering
    \includegraphics[width=\columnwidth]{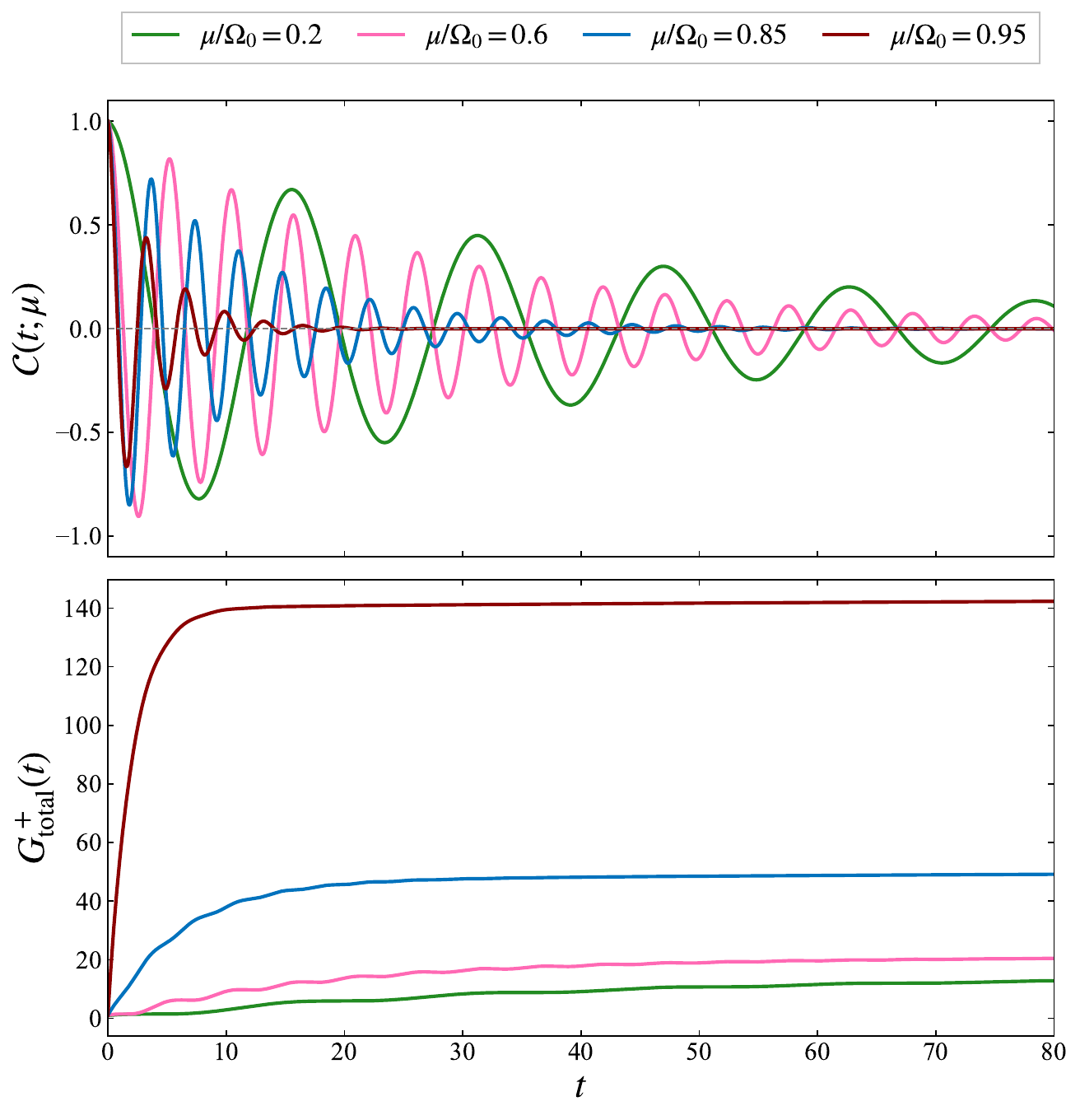}
    \refstepcounter{figure}\label{fig:cond}
    \smallskip
    \parbox{0.95\columnwidth}{\small\textbf{FIG.~\thefigure.} Illustrative normal-phase evolution at \(k=0\) for \(T=2.5\Omega_0\), \(\lambda=4\pi\), and \(\mu/\Omega_0=\{0.2,0.6,0.85,0.95\}\). The upper panel shows the normalized interference contrast \({\cal C}(t;\mu)\) of Eq.~\eqref{coherencecontrast}; for this illustrative dispersion, \(\Delta E=2\mu\), and the envelope uses \(\bar\Gamma_{\rm AS}=(\Gamma_{\rm AS}^+ +\Gamma_{\rm AS}^-)/2\). The lower panel uses \(E_-= \Omega_0-\mu\) and \(C_\beta=\coth[\beta E_-/2]\), so the late-time statistical amplitude grows as \(E_-\) approaches zero.}
\end{center}

Figure~\ref{fig:cond} then illustrates the normal-phase approach to the Bose-Einstein condensation threshold using the same benchmark damping scale in Eq.~\eqref{eq:aswidth}. The upper panel plots the normalized interference contrast in Eq.~\eqref{coherencecontrast}. For the illustrative \(k=0\) choice with common frequency \(\Omega_0\), the charge-sector separation is \(\Delta E=2\mu\), so larger \(\mu\) produces faster coherent oscillations and faster dephasing of the particle-antiparticle interference, while the benchmark damping suppresses the envelope. The lower panel shows the corresponding growth of the equal-time statistical amplitude as the shifted energy \(E_-=\Omega_0-\mu\) becomes small.\\

\begin{center}
    \centering
    \includegraphics[width=\columnwidth]{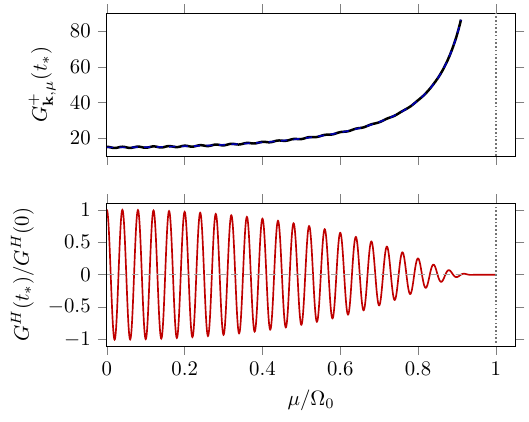}
    \refstepcounter{figure}\label{fig:gtvsmu}
    \smallskip
    \parbox{0.95\columnwidth}{\small\textbf{FIG.~\thefigure.} Chemical-potential dependence of the equal-time statistical correlator in the normal phase at \(t_\ast=80\Omega_0\), with \(T=7\Omega_0\), $\lambda=0.01$ and the benchmark damping scale of Eq.~\eqref{eq:aswidth}. The upper panel shows \(G_{\rm total}\), while the lower panel shows \(G^H(t_*)/G^H(0)\). The vertical dotted line marks the normal-phase endpoint \(\mu_c=\Omega_0\).}
\end{center}

Figure~\ref{fig:gtvsmu} gives a complementary scan in chemical potential. Instead of following the time evolution at fixed \(\mu\), it shows how the equal-time statistical correlator changes as the shifted pole energies move with \(\mu\). The upper panel compares the total correlator with the source-driven inhomogeneous contribution, making explicit that the growth near the threshold is driven by the same thermal factor that appears in Eq.~\eqref{thermalweights}. The lower panel shows that the normalized homogeneous remnant becomes subleading near the threshold.\\

The analysis above describes the normal phase of the charged scalar gas. In this regime the chemical potential shifts the particle and antiparticle sectors, but the field is not split into a condensate expectation value plus fluctuations and the lowest momentum mode remains part of the quasiparticle spectrum. The results should therefore be read as an approach-to-threshold calculation: they identify how the normal-phase propagators behave as the soft charge sector is driven toward Bose-Einstein condensation, but they do not attempt to describe the condensed phase itself.\\

The relevant quantities are the same shifted thermal weights that appear in the KMS step of Eq.~\eqref{shiftedkmsweights},
\begin{align}
  C_\beta^\pm
  &=
  \coth\!\left(\frac{\beta E_\pm}{2}\right)
  =
  1+2n_B(E_\pm),
  \nonumber\\
  n_B(E)&=\frac{1}{e^{\beta E}-1}.
  \label{thermalweights}
\end{align}
As long as these shifted energies remain positive, the thermal weights are finite and the nonequilibrium evolution is controlled by the damped quasiparticle modes appearing in the spectral function. Because the result is organized in terms of the shifted pole energies, it applies to either sign of \(\mu\) throughout the normal phase, as long as the quasiparticle description is meaningful.\\

Equation~\eqref{thermalweights} displays the normal-phase connection with Bose-Einstein condensation directly. If \(E_{\rm min}(\mu)\) denotes the smallest shifted quasiparticle energy,
\begin{equation}
  E_{\rm min}(\mu)=\min_{\bf k}\{E_+({\bf k},\mu),E_-({\bf k},\mu)\},
\end{equation}
then the normal-phase condition is \(E_{\rm min}(\mu)>0\). Equivalently, the soft branch is
\begin{equation}
  E_{\rm soft}({\bf k},\mu)=
  \begin{cases}
    E_-({\bf k},\mu)=\Omega_{\bf k}^{-}-\mu, & \mu>0,\\
    E_+({\bf k},\mu)=\Omega_{\bf k}^{+}+\mu, & \mu<0.
  \end{cases}
\end{equation}
At fixed temperature, increasing \(|\mu|\) lowers this branch until the lowest mode reaches the condensation threshold,
\begin{equation}
  E_{\rm min}(\mu_c)=0.
  \label{criticalcondition}
\end{equation}
For positive \(\mu\), this gives \(\mu_c\simeq\Omega_{{\bf k}=0}^{-}\) in the quasiparticle approximation, so the particle branch reaches the threshold first. For negative \(\mu\), the analogous condition is reached in the antiparticle branch.\\

Close to the threshold,
\begin{equation}
  n_B(E_{\rm min})\simeq \frac{T}{E_{\rm min}},
  \qquad
  C_\beta^{\rm min}\simeq \frac{2T}{E_{\rm min}},
  \label{becenhancement}
\end{equation}
and therefore the equal-time statistical propagator is enhanced by the same infrared factor that drives the growth of the zero-momentum occupation. This enhancement is the only BEC-related quantity computed here; no condensate fraction, critical exponent, or condensed-phase spectrum is extracted from the normal-phase propagator. In a finite volume this appears as a large but finite occupation of the lowest mode,
\begin{equation}
  n_0(\mu)=\frac{1}{\exp[\beta E_{\rm min}(\mu)]-1},
\end{equation}
while in the thermodynamic limit it signals the need to remove the zero mode from the fluctuation spectrum and assign the excess charge to a condensate density. In other words, the chemical potential is driven up to the lowest quasiparticle energy from below; once the threshold is reached, further charge is accommodated by the condensate rather than by pushing the normal-phase Bose distribution past its singular point.\\

Once the threshold is reached, the present quasiparticle treatment is no longer complete: the condensate must be treated as a macroscopic background field, and the propagator should be reorganized to include fluctuations around the condensate, possible anomalous correlators, and the mixing of amplitude and phase modes~\cite{Kapusta:1981aa,Haber:1981ts,Marko:2014hea,Nieves:2024viy,Nieves:2023yva,Nieves:2025pdi}. Thus the results presented here should be understood as describing the approach to the condensation threshold from the normal phase.\\

Together, Figs.~\ref{fig0}--\ref{fig:gtvsmu} show the progression from the coherent charge-sector transient to the full decohering response and then to the normal-phase Bose-Einstein-condensation precursor.

\section{\label{sec:conclusions}Conclusions}

We have analyzed a nonequilibrium coherent effect in charged scalar two-point functions at finite chemical potential. In frequency space, \(\mu\) acts through the expected shifted arguments of the retarded self-energy. In real time, the same shift separates the particle and antiparticle sectors and gives them independent phases, damping rates, and thermal weights. The calculation is controlled by an equilibrium-reservoir approximation and a narrow-width quasiparticle expansion; it does not include backreaction of the probe on the bath or the spectral reorganization of the condensed phase. Within this domain, the finite-density pole separation is kept explicitly rather than absorbed into a single overall phase.\\

The statistical propagator separates into two physically distinct pieces. The source-driven inhomogeneous part is fixed by the thermal bath and relaxes to the decoherent fluctuation-dissipation form. The homogeneous part carries initial-condition memory and contains the coherent particle-antiparticle interference generated by the finite chemical potential. This coherent signal is larger for larger initial correlations, but it is transient: damping suppresses the homogeneous piece, and dephasing from the shifted energy difference \(E_+-E_-\) washes out the interference before the late-time equilibrium limit is reached. The normalized contrast in Eq.~\eqref{coherencecontrast} makes this phase-sensitive part explicit. It should be understood as a diagnostic of the mixed charge-sector phases, not as a substitute for the complete homogeneous propagator.\\

Using the Aarts-Smit plasmon damping rate as a benchmark, we illustrated how larger chemical potential produces faster oscillations in the coherent sector while also accelerating the disappearance of the transient when the shifted damping rates increase. The same formalism provides a normal-phase diagnostic of the approach to Bose-Einstein condensation. As \(E_{\rm min}(\mu)\) approaches zero, the thermal factor in the statistical propagator behaves as \(C_\beta^{\rm min}\simeq2T/E_{\rm min}\), producing the infrared enhancement that drives the growth of the zero-momentum occupation. This is the normal-phase precursor of condensation, not a calculation of the condensate fraction, critical scaling, or the condensed-phase excitation spectrum; those require introducing the condensate as a macroscopic background field and reorganizing the propagators around it.\\

Extending the present real-time construction into the condensed phase is a
natural direction for future work. Once the chemical potential reaches the
condensation threshold, \(\mu_c\), the normal-phase quasiparticle description
breaks down and the scalar field must be separated into a macroscopic condensate
and fluctuations. In an interacting theory this reorganization generates
off-diagonal self-energies and anomalous statistical and spectral correlators,
so the Kadanoff-Baym equations must be enlarged beyond the normal propagator
sector. The resulting dynamics mixes the charge sectors and reorganizes the
spectrum into amplitude and phase modes, including the Goldstone mode associated
with spontaneous \(U(1)\) breaking~\cite{Kapusta:1981aa,Haber:1981ts,Marko:2014hea,Nieves:2023yva,Nieves:2024viy}.
Formulating the separation between transient finite-density dephasing and
decoherent reservoir relaxation in this broken-symmetry setting remains an
interesting open problem for nonequilibrium thermal field theory.

\section{\label{sec:acknowledgements}Acknowledgments}

A. C. acknowledges financial support from the Universidad Técnica Federico Santa María  through the Internal Master's Scholarship, Grant No. 003/2026.

\bibliography{biblio}

\end{document}